\documentclass[twocolumn,preprintnumbers,amsmath,amssymb,prd]{revtex4}
\usepackage[dvips]{graphicx}

\newcommand{\tr}{\mathrm{tr}}
\newcommand{\Pcal}{\mathcal{P}}
\newcommand{\xt}{\boldsymbol{x}_\perp}
\newcommand{\yt}{\boldsymbol{y}_\perp}
\newcommand{\pt}{p_t}
\newcommand{\del}{\boldsymbol{\partial}_\perp}
\newcommand{\GeV}{\;\text{GeV}}
\newcommand{\fm}{\;\text{fm}}
\newcommand{\GeVfm}{\;\text{GeV}\!\cdot\!\text{fm}^{-3}}
\newcommand{\LQCD}{\Lambda_{\text{QCD}}}
\newcommand{\half}{{\textstyle\frac{1}{2}}}
\newcommand{\quart}{{\textstyle\frac{1}{4}}}

\begin{document}
\title{Initial fields and instability in the classical model of the
  heavy-ion collision}
\author{Kenji Fukushima}
\affiliation{RIKEN BNL Research Center, Brookhaven National
 Laboratory, Upton, New York 11973, USA}
\begin{abstract}
 Color Glass Condensate (CGC) provides a classical description of
 dense gluon matter at high energies.  Using the McLerran-Venugopalan
 (MV) model we calculate the initial energy density
 $\varepsilon(\tau)$ in the early stage of the relativistic
 nucleus-nucleus collision.  Our analytical formula reproduces the
 quantitative results from lattice discretized simulations and leads
 to an estimate
 $\varepsilon(\tau\!=\!0.1\fm)=40\sim 50\GeVfm$ in the
 Au-Au collision at RHIC energy.  We then formulate instability with
 respect to soft fluctuations that violate boost invariance inherent
 in hard CGC backgrounds.  We find unstable modes arising, which is
 attributed to ensemble average over the initial CGC fields.
\end{abstract}
\preprint{RBRC-680}
\maketitle


  In the relativistic heavy-ion collision Color Glass Condensate (CGC)
describes the initial state of energetic gluon matter with the
transverse momentum $\pt$ up to the saturation scale $Q_s$ which
universally characterizes the hadron or nucleus wavefunction in the
small-$x$ regime~\cite{McLerran:1993ni,Kovner:1995ts,%
Kovchegov:1997ke,Krasnitz:1998ns,Krasnitz:2001qu,Lappi:2003bi,%
Lappi:2006fp,Lappi:2006hq,Romatschke:2005pm,Fries:2006pv}.  Given the
scale $Q_s$ at a certain value of Bjorken's $x$, the gluon
distribution probed by processes with $Q^2\ll Q_s^2$ is so dense that
coherent fields should be more relevant than the individual particle
picture during $\tau\lesssim Q_s^{-1}$.  Physically $Q_s^2$
corresponds to the transverse density of partons and is estimated by
the Golec-Biernat and W\"{u}sthoff fit,
$Q_s^2=Q_0^2(x_0/x)^\lambda A^{1/3}$, where $A$ is the atomic number.
We can expect $Q_s$ around $1\sim2\GeV$ for the Relativistic Heavy Ion
Collider (RHIC) and $2\sim3\GeV$ for the Large Hadron Collider (LHC)
in case of $A=197$ (Au-Au collision) assuming relevant $\pt$ is
$\sim 1\GeV$.  This transient but still coherent gluon matter, which
is often referred to as ``Glasma''~\cite{Lappi:2006fp}, should melt
toward a quark-gluon plasma.

  The physical property of Glasma has been mainly analyzed by
numerical simulations in the lattice discretized
formulation~\cite{Krasnitz:1998ns,Krasnitz:2001qu,Lappi:2003bi,%
Lappi:2006fp,Lappi:2006hq,Romatschke:2005pm}.  In this paper we aim
to approach Glasma in an analytical way along a similar line to the
near-field expansion proposed by Fries-Kapusta-Li~\cite{Fries:2006pv}.
The analytical method is desirable for a deeper insight into the
Glasma, which presumably exists up to
$\tau\lesssim Q_s^{-1}\sim 0.1\fm$ in the Au-Au (central) collision at
RHIC energy, $\sqrt{s}=200\GeV/\text{nucleon}$, or even longer
depending on the interpretation of the Glasma.  In particular, the
problem of \textit{early thermalization} still has interesting
unanswered questions~\cite{Kovchegov:2005ss}.  We will specifically
address the following; is there any unstable mode growing around the
initial CGC fields right after the collision?   If any, it could speed
up thermalization (or isotropization) even in the classical regime
($\tau\lesssim Q_s^{-1}$), besides non-Abelian plasma
instabilities~\cite{Mrowczynski:1988dz,Mrowczynski:2004kv,%
Arnold:2003rq,Romatschke:2006wg} which take place at later times.  The
pioneering numerical simulation~\cite{Romatschke:2005pm} suggests the
existence of ``Glasma instability'', though the literal time scale of
instability seems to be greater than $Q_s^{-1}$ by three order of
magnitude, probably because of the choice of tiny instability seeds.
The delay in the instability onset has also been pointed out in the
Hard Expanding Loop (HEL) approach to non-Abelian plasma
instabilities~\cite{Romatschke:2006wg}.


  We shall start with the boost-invariant CGC solution and estimate
an initial energy density.  It is convenient to adopt the Bjorken
coordinates spanned by the proper time $\tau=\sqrt{t^2-z^2}$ and the
space-time rapidity $\eta=\half\ln[(t+z)/(t-z)]$.  The radial gauge
$A_\tau=0$ is understood throughout this work.  The canonical momenta
(chromo-electric fields) are read in this gauge as
\begin{equation}
 E^i = \tau\partial_\tau A_i \;,\quad
 E^\eta = \tau^{-1}\partial_\tau A_\eta \;.
\label{eq:momentum}
\end{equation}
It should be mentioned that the metric is $g_{\tau\tau}=1$,
$g_{\eta\eta}=-\tau^2$, and $g_{xx}=g_{yy}=-1$ in accord with the
convention in Refs.~\cite{Romatschke:2005pm,Fukushima:2006ax}.  The
equations of motion derived from Hamilton's equations lead us to
\begin{equation}
 \partial_\tau E^i = \tau^{-1} D_\eta F_{\eta i}
  +\tau D_j F_{ji} \;,\quad
 \partial_\tau E^\eta = \tau^{-1} D_j F_{j\eta} \;.
\label{eq:eom}
\end{equation}
These are the basic equations for the classical description valid in
the early stage right after the collision.  The initial condition is
uniquely determined by boundary matching at singularities of the color
sources $\rho^{(1)}(\xt)\delta(x^-)$ and $\rho^{(2)}(\xt)\delta(x^+)$
representing the propagation of Lorentz contracted
nuclei~\cite{Kovner:1995ts,Fukushima:2006ax}, as follows;
\begin{equation}
 \begin{split}
 & A_{i(0)} = \alpha^{(1)}_i + \alpha^{(2)}_i \;,\quad
   A_{\eta(0)} = 0 \,,\\
 & E^i_{(0)} = 0 \;,\quad
   E^\eta_{(0)} = ig\bigl[\alpha^{(1)}_i,\alpha^{(2)}_i\bigr] \,,
 \end{split}
\label{eq:initial}
\end{equation}
where $\alpha^{(1)}_i$ and $\alpha^{(2)}_i$ are the gauge fields at
$\tau<0$ associated with the right-moving nucleus along the $x^+$ axis
and the left-moving nucleus along the $x^-$
axis~\cite{Kovchegov:1996ty}.  It takes a pure-gauge form,
$\alpha_i(\xt) = -(1/ig)V(\xt)\partial_i V^\dagger(\xt)$, with the
Wilson line defined by
\begin{equation}
 V^\dagger(\xt) = \Pcal\exp\biggl[-ig\int d z^-\,
  \frac{1}{\del^2}\rho^{(1)}(\xt)\delta(z^-) \biggr] \;,
\end{equation}
for $\alpha^{(1)}_i$.  The Wilson line for $\alpha^{(2)}_i$ is given
by replacement of $x^-$ and $\rho^{(1)}(\xt)$ by $x^+$ and
$\rho^{(2)}(\xt)$ in the above expression.  We can compute the
expectation value of physical observables by means of the average over
the random color distribution inside nuclei using
\begin{equation}
 \bigl\langle\rho_a^{(m)}(\xt)\rho_b^{(n)}(\yt)\bigr\rangle
  = g^2\mu^2\,\delta^{mn}\,\delta_{ab}\,\delta^{(2)}(\xt-\yt) \;.
\label{eq:Gaussian}
\end{equation}
Here $\mu$ is the only dimensionful scale in the McLerran-Venugopalan
(MV) model and related to the saturation scale $Q_s$.  We will later
present all dimensionful quantities in unit of $\mu$.


  Let us evaluate the initial energy density of the
fields~(\ref{eq:initial}) at $\tau=0$ with the color source
average~(\ref{eq:Gaussian}).  To do this, we need to take an average
of four Wilson lines
$\sim\langle V(\xt)V^\dagger(\boldsymbol{y}_\perp)
V(\boldsymbol{u}_\perp)V^\dagger(\boldsymbol{v}_\perp)\rangle$.  We
can find an algebraic technique in the appendix of
Ref.~\cite{Blaizot:2004wv} and it is even possible to write a formal
expression down for more generic color
structure~\cite{Fukushima:2007dy}.  Alter all, it turns out that the
transverse fields are vanishing and the longitudinal chromo-magnetic
fields, $B^\eta_{(0)}=F_{12(0)}$, are~\cite{Lappi:2006hq}
\begin{equation}
 \frac{g^2}{(g^2\mu)^4}\cdot \Bigl\langle 2\tr\bigl(B^\eta_{(0)}
  \bigr)^2\Bigr\rangle
 = \frac{1}{32}N_c (N_c^2-1)\, \sigma^2 \;.
\label{eq:mag_l}
\end{equation}
The number of color is $N_c=3$ in QCD.\ \ We defined $\sigma$
resulting from the two-point function in terms of $\alpha^{(m)}_i$.
In order to make a direct comparison to the numerical simulation
transparently, we shall make use of the lattice regularization, which
gives
\begin{align}
 \sigma &= \frac{1}{2L^2}\!\sum_{n_i=1-L/2a}^{L/2a}\!
  \frac{1}{2-\cos(2\pi n_1 a/L)-\cos(2\pi n_2 a/L)} \notag\\
 &\simeq \frac{1}{2\pi}\ln(c L/a ) \;.
\label{eq:log}
\end{align}
Here $L$ is the size of the system fixed by $L^2=\pi R_A^2$, and $a$
is the lattice spacing.  We got rid of the zero-mode $n_1=n_2=0$
because of global neutrality.  We numerically checked that the above
logarithmic form with adjusted by a constant $c\simeq1.36$ is a quite
good approximation.  Some further calculations end up with the same
amount of the chromo-electric field squared;
$\langle2\tr(E^\eta_{(0)})^2\rangle = \langle 2\tr(B^\eta_{(0)})^2\rangle$.
As a result, we can estimate the initial energy density as
\begin{equation}
 \frac{g^2}{(g^2\mu)^4}\cdot \varepsilon_{(0)} = \frac{3}{4}\sigma^2
\label{eq:energy_zero}
\end{equation}
with $N_c=3$ substituted.  This $a$ and $L$ dependent result should be
interpreted carefully, while the quantitative output somehow agrees
with the latest simulation by Lappi;  our estimate by
Eqs.~(\ref{eq:log}) and (\ref{eq:energy_zero}) yields 0.81 and 0.90
for $L/a=500$ and $700$ which are close to 0.76 and 0.88 reported in
Ref.~\cite{Lappi:2006hq}.  The logarithmic singularity has been found
also in Refs.~\cite{Lappi:2006hq,Fries:2006pv}.  The singularity
arises from the approximations that we regarded the colliding nuclei
as infinitely thin in the longitudinal direction and that the random
color distribution is uncorrelated at arbitrary microscopic scale in
transverse space.

  Next, we will step away from singularity located at $\tau=0$ by the
near-field expansion in terms of $\tau$, i.e.,
$\mathcal{O}=\mathcal{O}_{(0)}+\mathcal{O}_{(1)}\tau+\mathcal{O}_{(2)}\tau^2+\cdots$.
The first-order corrections are vanishing, and the second-order fields
are
\begin{equation}
 \begin{split}
 & A_{i(2)} = \half E^i_{(2)} = \quart D_{j(0)}F_{ji(0)} \;,\\
 & A_{\eta(2)} = \half E^\eta_{(0)} \;,\quad
   E^\eta_{(2)} = \half D_{j(0)}F_{j\eta(2)} \;,
 \end{split}
\end{equation}
where
\begin{equation}
 \begin{split}
 F_{ji(0)} &= -ig\bigl( \bigl[\alpha^{(1)}_j,\alpha^{(2)}_i\bigr]
  +\bigl[\alpha^{(2)}_j,\alpha^{(1)}_i\bigr] \bigr) \;,\\
 F_{j\eta(2)} &= \half D_{j(0)}E^\eta_{(0)} \;,
 \end{split}
\end{equation}
which physically represent the initial longitudinal and second-order
transverse chromo-magnetic fields.

  Using these expressions we calculate the contributions to the energy
density of order $\tau^2$ to find the same amount of chromo-magnetic
and chromo-electric fields again.  Since the initial state has
non-zero longitudinal fields, it follows that the cross terms between
the zeroth and second-order terms give
\begin{align}
 & \frac{g^2}{(g^2\mu)^4}\cdot 2\Bigl\langle2\tr\bigl(B^\eta_{(2)}
  B^\eta_{(0)}\bigr)\Bigr\rangle = \frac{g^2}{(g^2\mu)^4}\cdot 2\Bigl
  \langle2\tr\bigl(E^\eta_{(2)}E^\eta_{(0)}\bigr)\Bigr\rangle \notag\\
 & = -\frac{1}{32}N_c(N_c^2-1)\, \sigma\cdot\chi
  +\mathcal{O}(\sigma^3)\;,
\end{align}
where we defined
\begin{equation}
 \chi = \frac{1}{L^2}\sum_{n_i=1-L/2a}^{L/2a} \simeq \frac{1}{a^2} \;.
\label{eq:chi}
\end{equation}
We dropped terms proportional to $\sigma^3$ not containing $\chi$
because $\chi\gg\sigma$ when $a$ is small.  In the same approximation
the transverse fields of order $\tau^4$ (that is, $\tau^2$-order in
the energy density) result in
\begin{align}
 & \frac{g^2}{(g^2\mu)^4}\cdot\Bigl\langle2\tr\bigl(B^i_{(2)}
  B^i_{(2)}\bigr)\Bigr\rangle = \frac{g^2}{(g^2\mu)^4}\cdot
  \Bigl\langle2\tr\bigl(E^i_{(2)}E^i_{(2)}\bigr)\Bigr\rangle \notag\\
 &= \frac{1}{64}N_c(N_c^2-1)\, \sigma\cdot\chi +\mathcal{O}(\sigma^3)\;.
\end{align}

\begin{figure}
 \includegraphics[width=7cm]{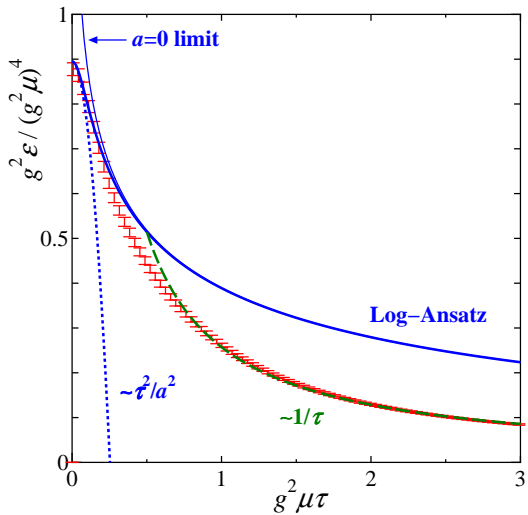}
 \caption{Comparison of the energy density in case of $L/a=700$; the
 data with error bar is taken from Ref.~\cite{Lappi:2006hq}.  The
 dotted and solid curves represent the naive expansion in
 Eq.~(\ref{eq:energy_expand}) and the log-ansatz in
 Eq.~(\ref{eq:log_ansatz}), respectively.  The dashed curve scales as
 $1/\tau$ whose starting point is chosen at $g^2\mu\tau=0.5$, meaning
 that the ``formation time''~\cite{Krasnitz:1998ns} being
 $g^2\mu\tau_D\sim0.5$.}
 \label{fig:energy}
\end{figure}

  After all, we get the expanded series,
\begin{align}
 \frac{g^2}{(g^2\mu)^4}\cdot\varepsilon &\simeq
  \frac{g^2}{(g^2\mu)^4}\Bigl[\varepsilon_{(0)}
  + \varepsilon_{(2)} \tau^2 \Bigr] \notag\\
 &= \frac{1}{32} N_c (N_c^2-1) \sigma \biggl[ \sigma - \pi
  \frac{(g^2\mu\tau)^2}{(g^2\mu a)^2} \biggr] \;.
\label{eq:energy_expand}
\end{align}
It is obvious from Eq.~(\ref{eq:energy_expand}) that the $\tau$
expansion behaves badly for small value of $a$, which is also clear by
the dotted curve in Fig.~\ref{fig:energy} that plots
Eq.~(\ref{eq:energy_expand}).

  The naive $\tau$ expansion is, in fact, ill-defined.  It is because,
as pointed out in Ref.~\cite{Lappi:2006fp}, the energy density behaves
as $\sim(\ln\tau)^2$ near $\tau=0$ when the colliding nuclei are
infinitely thin.  Therefore, the naive Taylor expansion around
$\tau=0$ is meaningless.  Nevertheless, we stress that we can derive
meaningful information from Eq.~(\ref{eq:energy_expand}); we know that
the asymptotic form $\sim(\ln\tau)^2$ in the $a\to0$ limit and we
also know that the regularized expansion $\sim c_1\ln(L/a)[\ln(L/a)
  +c_2(\tau/a)^2]$ with $a$ kept finite.  The simplest analytical
function satisfying these two requirements is
$\sim c_1\{\ln[L^2/(a^2-c_2 \tau^2)]\}^2$, that means,
\begin{equation}
 \frac{g^2}{(g^2\mu)^4}\cdot\varepsilon \simeq \frac{3}{4}\Biggl\{
  \frac{1}{4\pi}\ln\biggl[
  \frac{c^2(g^2\mu L)^2}{(g^2\mu a)^2 + \pi (g^2\mu\tau)^2}
  \biggr]\Biggr\}^2 \;.
\label{eq:log_ansatz}
\end{equation}
The comparison to data obtained in the numerical simulation is
presented in Fig.~\ref{fig:energy}.  This simple log-ansatz works well
as long as $g^2\mu\tau\lesssim0.5$ and is stable under the $a\to0$
limit as shown by a thin curve in the figure.

  So far, we reached an ansatz~(\ref{eq:log_ansatz}) with the infrared
cut-off provided by the nucleus size $L$ in a heuristic way.  In
reality, however, the long-ranged correlation should be cut off by the
confining scale $\sim\LQCD^{-1}$ rather than $L$.  In the continuum
limit, hence, the initial energy density in the central collision at
$\tau\ll(g^2\mu)^{-1}$ should be estimated by
\begin{equation}
 \varepsilon = \frac{3}{16\pi^2 g^2}(g^2\mu)^4 \Bigl\{
  \ln(\LQCD^{-1}/\tau) \Bigr\}^2 \;.
\label{eq:energy}
\end{equation}
In writing Eq.~(\ref{eq:energy}) we put the constants $c$ and $\pi$
appearing in Eq.~(\ref{eq:log_ansatz}) away into ambiguity of
$\LQCD$.  We remark that the $\LQCD$-dependence would be milder than
the above in the regime after the ``formation time'' as investigated
in Ref.~\cite{Krasnitz:1998ns}.

  It is interesting to apply our formula (\ref{eq:energy}) to the
Au-Au collision at RHIC in the physical unit.  We make use of the
parameter choice as commonly used in the numerical simulation, i.e.,
$g^2/4\pi=1/\pi$ and
$g^2\mu=2\GeV$~\cite{Krasnitz:1998ns,Krasnitz:2001qu,Lappi:2003bi,Fries:2006pv}.
As for the confining scale, we vary $\LQCD^{-1}$ from $1\fm$ to
$12\fm\simeq L$.  The results are summarized as follows;
\begin{center}
 \begin{tabular}{|c||c|c|c|c|c|c|}
\hline
$\LQCD^{-1}$ [fm] & 1 & 3 & 5 & 8 & 10 & 12\\
\hline
$\varepsilon(\tau\!=\!0.1\fm)$
  [$\GeV\!\!\cdot\!\!\fm^{-3}$] & 53 & 115 & 152 & 191 & 211 & 228\\
\hline
corrected
  [$\GeV\!\!\cdot\!\!\fm^{-3}$] & 36 & 77 & 102 & 128 & 142 & 153\\
\hline
 \end{tabular}
\end{center}
Our log-ansatz overestimates the energy density and the third row
shows the corrected values with a factor 0.67 inferred from
Fig.~\ref{fig:energy}.  This factor might depend on $\LQCD^{-1}$, and
thus, the numbers listed in the second and third rows should be
considered as the upper and lower bounds.

  It is a natural choice to take the confining scale as the nucleon
size $\sim 1\fm$, and the estimate of the initial energy density is
then $\varepsilon(\tau\!=\!0.1\fm)=40\sim50\GeV\!\cdot\!\fm^{-3}$.
This value is significantly smaller than the previous estimates,
$130\GeV\!\cdot\!\fm^{-3}$ in Ref.~\cite{Lappi:2006hq} and
$260\GeV\!\cdot\!\fm^{-3}$ in Ref.~\cite{Fries:2006pv}, reflecting
difference between the choices $\LQCD^{-1}=1\fm$ and
$\LQCD^{-1}=L\sim12\fm$, but rather consistent with the simulation with
color neutrality of finite nuclei taken into
account~\cite{Krasnitz:2001qu} that found
$\epsilon(\tau\!=\!\tau_D\!\simeq\!0.3\fm)
=7.1\sim40\GeV\!\cdot\!\fm^{-3}$.

  When $g^2\mu\tau$ becomes larger, the energy density comes to scale
as $\sim\tau_0/\tau$ because of (almost free streaming) longitudinal
expansion~\cite{Romatschke:2005pm,Kovchegov:2005ss}.  It should be
noted that the scaling law in the classical regime is different from
the (one-dimensional) hydrodynamic one $\sim(\tau_0/\tau)^{4/3}$.  For
reference we plot the scaling behavior
$\varepsilon(\tau)/\varepsilon(\tau_0)=\tau_0/\tau$ in
Fig.~\ref{fig:energy} indicated by the dashed curve with a choice of
$g^2\mu\tau_0=0.5$.  The expanding system at late times is dilute so
that this scaling behavior is to be justified by the solution of
Eq.~(\ref{eq:eom}) in the weak field limit, which in fact scales
as~\cite{Kovner:1995ts,Lappi:2006hq}
\begin{equation}
 A_i \sim A_\eta \sim 1/\sqrt{\tau} \;.
\label{eq:asymptotic}
\end{equation}
The energy density is dominated only by the Abelian part
$\sim(\partial A)^2$, hence it follows the $\tau_0/\tau$ scaling.  We
would comment on a curious observation that, if we extrapolate our
estimate $\varepsilon(\tau\!=\!0.1\fm)$ up to $\tau=1\fm$ assuming the
$\tau_0/\tau$ scaling, the initial energy density obtained
accordingly is very close to the standard estimate by means of the
Bjorken formula;
$\epsilon(\tau\!=\!1\fm)\sim 5.1\GeV\!\cdot\!\fm^{-3}$.
\\


  We shall next consider the problem of instability in the rest of
this paper.  We treat fluctuations $\delta A_i$ and $\delta E^i$ in
the linear order around the boost-invariant CGC background which we
discussed above.  As formulated in Ref.~\cite{Fukushima:2006ax},
$\delta E^\eta$ should be constrained by the Gauss law and we drop
$\delta A_\eta$ because it is accompanied by $\tau^2$ from the
metric.  In what follows we regard $\eta$-dependent fluctuations,
$\delta A_i$ and $\delta E^i$, as the ``soft'' fields and
$\eta$-independent CGC fields as the ``hard'' background which brings
about instability.  The linearized equations of motion are
\begin{equation}
 \begin{split}
 \tau\partial_\tau\delta \tilde{A}_i &= \delta \tilde{E}^i \;,\\
 \partial_\tau \delta \tilde{E}^i &= -\tau^{-1} \nu^2
  \delta \tilde{A}_i + \tau\, G^{-1}_{ij}\delta \tilde{A}_j \;,
 \end{split}
\label{eq:diff}
\end{equation}
where we introduced the Fourier transform $\delta\tilde{A}_i$ from
$\eta$ to the wave number $\nu$ (i.e.\
$\partial^2_\eta\delta A_i(\eta)\to -\nu^2\delta\tilde{A}_i(\nu)$) and
we denoted the inverse of the transverse background gluon propagator
as $G^{-1}_{ij}$ whose definition is
\begin{equation}
 G^{-1ab}_{ij} = \delta_{ij}(D_k D_k)^{ab} - (D_i D_j)^{ab}
  +2g f^{acb}F_{ij}^c \;.
\end{equation}
We note that, in correspondence to the HEL approach, the color current
encoding the anisotropic distribution of hard background is identified
as $j^a_i =\bigl[ G^{-1ab}_{ij}-\delta^{ab}\bigl( \delta_{ij}
\partial_k\partial_k - \partial_i\partial_j \bigr) \bigr]
\delta\tilde{A}_j^b$.
Although it is not clear whether this current could have anything to
do with that in the HEL approach after the ensemble average, we can
shortly confirm that instability may occur even in the purely
classical regime.

  It is easy to solve Eq.~(\ref{eq:diff}) to obtain
$\delta\tilde{A}_i$ as a function of a given constant CGC background,
$G^{-1}_{(0)}$ at initial time, because we can diagonalize
$G^{-1}_{(0)}$ in a proper basis of $\delta\tilde{A}_i$.  If we write
its eigenvalues as $\lambda$, the solution is
\begin{equation}
 \delta\tilde{A}_i(\lambda) = c_{1i}\,\text{Re}
  I_{i\nu}(\sqrt{\lambda}\tau)
  +c_{2i}\,\text{Im}I_{i\nu}(\sqrt{\lambda}\tau) \,,
\label{eq:unstable}
\end{equation}
for $\lambda>0$ (which exponentially grows) and
\begin{equation}
 \delta\tilde{A}_i(\lambda) = c_{1i}\,\text{Re}
  J_{i\nu}(\sqrt{|\lambda|}\tau)
  +c_{2i}\,\text{Im}J_{i\nu}(\sqrt{|\lambda|}\tau) \,,
\label{eq:stable}
\end{equation}
for $\lambda<0$ (which oscillatorily diminishes), where $J_n(x)$ and
$I_n(x)$ are the first-kind and modified Bessel functions.  These
special functions are singular as
$(\sqrt{|\lambda|}\tau)^{\pm i\nu}$ which furiously rotates in complex
space as $\tau\to0$.  At later time when the asymptotic
behavior~(\ref{eq:asymptotic}) realizes due to expansion, $\lambda$ is
no longer a constant but a function of time like $\sim \xi/\tau$ with
some dimensionful constant $\xi$, which
results in the solutions
$I_{2\pm i\nu}(2\sqrt{\xi\tau})$ for $\xi>0$ and
$J_{2\pm i\nu}(2\sqrt{|\xi|\tau})$ for $\xi<0$.  For other general
cases, the $\tau$-dependence in the eigenvalue is intricate, and one
needs to solve Eq.~(\ref{eq:diff}) numerically.

\begin{figure}
 \includegraphics[width=7cm]{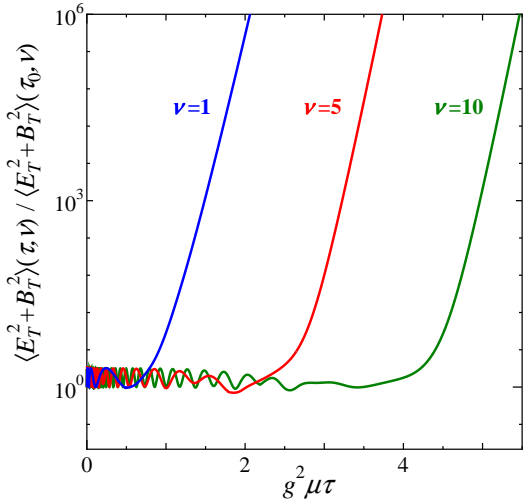}
 \caption{Instability tendency with the initial CGC background fixed
 at $\tau=0$ in case of $L/a=700$ for different wave numbers $\nu=1$,
 $5$, and $10$ from the left to the right.}
 \label{fig:inst}
\end{figure}

  We do not treat such a general case but what we will pursue here is
to clarify whether the \textit{initial} CGC fields could induce
exponential growth for soft degrees of freedom.  We consider this
because the initial CGC configurations at $\tau=0$ are the most
(spatially) anisotropic (namely, large longitudinal and zero
transverse fields) and thus anticipated to cause the most unstable
modes.  Under such extreme circumstances we expect that the physics of
instability becomes clear.  Thus, as a first trial, it should be an
appropriate starting point.  We can say that what we will do is to
extract the \textit{tendency} toward instability under the presence of
the CGC background.  For simplicity we will focus on the case that
fluctuations are uniform in transverse space, i.e.\
$\partial_i\delta\tilde{A}=0$, which should be the most unstable, as
adopted in Ref.~\cite{Romatschke:2006wg}.

  We should calculate the Gaussian average~(\ref{eq:Gaussian}) of
the transverse chromo-magnetic field
$\langle(B_i)^2\rangle\simeq\langle(\nu\delta\tilde{A}_i)^2\rangle$
and the chromo-electric field
$\langle(E_i)^2\rangle\simeq\langle(\tau\partial_\tau\delta\tilde{A}_i)^2\rangle$
which were as small as $\sim\tau^4$ previously but non-zero this time
with fluctuations depending on $\eta$, contributing to the
longitudinal pressure~\cite{Romatschke:2005pm}.  The straightforward
calculation is, however, technically hard.  We will approximately do
it by picking the mean value up,
\begin{equation}
 \bigl\langle G_{(0)ij}^{-1ab}\bigr\rangle = \delta_{ij}\delta^{ab}
  \bar{\lambda} = -\delta_{ij}\delta^{ab}\frac{3}{8}\sigma\,
  (g^2\mu)^2 \;,
\label{eq:mean}
\end{equation}
where $\sigma$ is defined in Eq.~(\ref{eq:log}), and taking the
ensemble average over its dispersion,
\begin{equation}
 \bigl\langle G_{(0)ik}^{-1ac}G_{(0)kj}^{-1cb}\bigr\rangle
  -\delta_{ij}\delta^{ab}\bar{\lambda}^2
 =\delta_{ij}\delta^{ab}\delta\lambda^2 
 = \delta_{ij}\delta^{ab}\frac{3}{8}\chi\,(g^2\mu)^2 ,
\label{eq:dispersion}
\end{equation}
where $\chi$ is defined in Eq.~(\ref{eq:chi}).

  Because $\bar{\lambda}$ is negative, the soft fluctuations in the
vicinity of the averaged CGC background are stable belonging to the
type of solution~(\ref{eq:stable}).  The eigenvalue of $G_{(0)}^{-1}$
distributes according to random CGC configurations and spreads from
$\bar{\lambda}$ with the dispersion $\delta\lambda$, meaning that some
CGC configurations may have negative $\lambda$.  That is, if we
evaluate,
\begin{equation}
 \bigl\langle\mathcal{O}[\delta\tilde{A}(\lambda)]\bigr\rangle
  \simeq \int_{-\infty}^\infty\!\!d\lambda\,
  \mathcal{O}[\delta\tilde{A}(\lambda)]\,
  e^{-(\lambda-\bar{\lambda})^2/2\delta\lambda^2} \,,
\end{equation}
using Eq.~(\ref{eq:unstable}), the contributions near
$\lambda\simeq\bar{\lambda}$ dominate only when time is small until
the negative $\lambda$ constituents grow up as time elapses.  The
transverse field strengths obtained in this way are plotted in
Fig.~\ref{fig:inst}.

  To draw Fig.~\ref{fig:inst}, we chose the initial time
$g^2\mu\tau_0=0.001$ at which we set $c_1$ and $c_2$ of
Eq.~(\ref{eq:unstable}) or (\ref{eq:stable}) by the initial condition,
$\delta\tilde{A}_i=c/\sqrt{\nu}$ and
$\delta\tilde{E}_i=\tau_0\partial_\tau\delta\tilde{A}_i=c\sqrt{\nu}$,
inspired by quantum fluctuations discussed in
Ref.~\cite{Fukushima:2006ax}.  It is interesting to see that this
specific initial condition
($\delta\tilde{E}_i \sim \nu\delta\tilde{A}_i$) makes
$\langle(B_i)^2\rangle\simeq\langle(\nu\delta\tilde{A}_i)^2\rangle$
and $\langle(E_i)^2\rangle$ comparable to each other, leading to their
almost alternate oscillations.  That is why the sum of transverse
field strengths depicted in Fig.~\ref{fig:inst} never come close to
zero, which makes a contrast to the results in
Ref.~\cite{Romatschke:2006wg}.

  We can conclude that there is certainly the tendency toward
instability associated with initial CGC background.  The onset of
instability in the present case is located much earlier than preceding
works.  It is because we only investigated the strongest instability
encompassed in the initial CGC fields.  Because the CGC background
itself evolves with time, in fact, the genuine growth of instability
should be slower and weaker than shown in Fig.~\ref{fig:inst}.
Nevertheless, we can learn the qualitative character of the ``Glasma''
instability.  The intuitive picture is as follows.  The soft
fluctuations of gluon fields are non-Abelian charged and feel a force
under influence from the CGC background.  The ensemble of random CGC
distribution contains not only color fields which suppress the color
current provided by charged soft fluctuations but also color fields
which amplify the current.  Although the current is suppressed on
average, the large $\tau$ behavior is predominantly determined by
mixture of CGC fields which enhance the input current.  Therefore, we
think that it is rare fluctuation in the CGC ensemble from which the
Glasma instability can occur.

  It is necessary to deal with $\lambda(\tau)$ as not a constant but a
function of $\tau$ in order to quantify instability further.  In our
treatment mentioned above we dropped the effect of longitudinal
expansion for the hard part, while the exponential growth should take
a form of
$I_{2\pm i\nu}(2\sqrt{\xi\tau})\sim \tau^{-1/2}\exp[2\sqrt{\xi\tau}]$
asymptotically when $\lambda\sim\xi/\tau$, as we remarked before.  The
analytical estimate of $\xi$ deserves future clarification.  Also, we
have to evaluate $\bar{\lambda}$ and $\delta\lambda$ in a resummed
form like Eq.~(\ref{eq:log_ansatz}) beyond the naive expressions
(\ref{eq:mean}) and (\ref{eq:dispersion}).  As a matter of fact, the
growth rate seems to be determined by $\bar{\lambda}$ and
$\delta\lambda$ regardless of $\nu$ in view of our results in
Fig.~\ref{fig:inst}.  Quantitative details of an analytical
description should be improved with guided by systematic instability
studies in the numerical simulation in the future.

  In summary, we developed an analytic formula to estimate the initial
energy density.  Our conclusion is
$\epsilon(\tau=0.1\fm)=40\sim50\GeV\!\cdot\!\fm^{-3}$ in the (central)
Au-Au collision at RHIC.\ \ The uncertainty comes from the infrared
cut-off (or confining) scale.  Also, we analyzed the tendency toward
instability in the presence of the initial CGC background fixed at
$\tau=0$.  We found that there exist unstable modes as a result of the
ensemble average of random CGC configurations, some of which
strengthen the color current brought in by soft fluctuations.
Although the Glasma instability might have a connection to non-Abelian
plasma instabilities at a deeper level, we would emphasize that we
could understand the Glasma instability not relying on the picture of
plasma instabilities that are premised on anisotropic momentum
distribution of hard particles.  The bottom line is, thus, that the
Glasma instability exists from $\tau=0$ even when the particle picture
is irrelevant yet.

  The author thanks Larry McLerran, Raju Venugopalan, Tuomas Lappi,
Kazu Itakura, and Yuri Kovchegov for useful discussions.  He is
especially grateful to Tuomas for sending him the simulation data and
to Raju for helpful comments to the manuscript.  This work was
supported by RIKEN BNL Research Center and the U.S.\ Department of
Energy under cooperative research agreement \#DE-AC02-98CH10886.

\end{document}